\documentclass[twocolumn,superscriptaddress]{revtex4-1}
\usepackage{epsfig,graphicx}
\usepackage{dcolumn}
\usepackage{bm}
\usepackage{amsmath}
\usepackage{epsfig}
\usepackage{graphics,graphicx,dcolumn,bm,fleqn,epic,eepic,float}
\usepackage{amssymb,amsmath,multirow,rotate,color}
\usepackage{times}
\usepackage{verbatim}
\usepackage{subfigure}
\usepackage{color}
\definecolor{red}{rgb}{1,0,0}

\definecolor{blue}{rgb}{0,0,1}

\begin{document}

\title{Evolutionary hypergame dynamics}

\author{JunJie Jiang}
\affiliation{School of Electrical, Computer and Energy Engineering,
Arizona State University, Tempe, AZ 85287, USA}

\author{Yu-Zhong Chen}
\affiliation{School of Electrical, Computer and Energy Engineering,
Arizona State University, Tempe, AZ 85287, USA}

\author{Zi-Gang Huang}
\affiliation{The Key Laboratory of Biomedical Information Engineering of 
Ministry of Education, Institute of Biomedical engineering, School of Life 
Science and Technology, Xi'an Jiaotong University, Xi'an, 710049, China}

\author{Ying-Cheng Lai} \email{Ying-Cheng.Lai@asu.edu}
\affiliation{School of Electrical, Computer and Energy Engineering,
Arizona State University, Tempe, AZ 85287, USA}
\affiliation{Department of Physics, Arizona State University,
Tempe, Arizona 85287, USA}

\date{\today}

\begin{abstract}

A common assumption employed in most previous works on evolutionary game 
dynamics is that every individual player has full 
knowledge about and full access to the complete set of available strategies. 
In realistic social, economical, and political systems, diversity in the 
knowledge, experience, and background among the individuals can be expected. 
Games in which the players do not have an identical strategy set are 
hypergames. Studies of hypergame dynamics have been scarce, especially those
on networks. We investigate evolutionary hypergame dynamics on regular 
lattices using a prototypical model of three available strategies, in which 
the strategy set of each player contains two of the three strategies. 
Our computations reveal that more complex dynamical phases emerge from
the system than those from the traditional evolutionary game dynamics
with full knowledge of the complete set of available strategies, which
include single-strategy absorption phases, a cyclic competition
(``rock-paper-scissors'') type of phase, and an uncertain phase in which
the dominant strategy adopted by the population is unpredictable.
Exploiting the pair interaction and mean field approximations, we 
obtain a qualitative understanding of the emergence of the single strategy 
and uncertain phases. We find the striking phenomenon of strategy revival 
associated with the cyclic competition phase and provide a qualitative 
explanation.Our work demonstrates that the diversity in the individuals' 
strategy set can play an important role in the evolution of strategy 
distribution in the system. From the point of view of control, the emergence 
of the complex phases offers the possibility for harnessing evolutionary game 
dynamics through small changes in individuals' probability of strategy 
adoption.

\end{abstract}

\maketitle

\section{Introduction} \label{sec:intro}

Evolutionary games are a powerful mathematical and computational paradigm
to gain qualitative and quantitative insights into a variety of phenomena
in diverse disciplines such as biology, ecology, economics, social and 
political sciences~\cite{SB:2016,Smith:book,VM:book,Colman:book}. A key success
of evolutionary game theory is the discovery of the general principles that 
govern the emergence and evolution of cooperation, a phenomenon that seems to 
contradict the principle of natural selection, which provides a convincing 
explanation of its ubiquity in both animal world and human 
society~\cite{SP:1973,HA:1981,Dugatkin:1997,SMWB:2004,Nowak:2006,
VGS:2006,SS:2007}. 
In particular, in most previous works on evolutionary game dynamics, 
an assumption is that each and every player participating in the game has full 
knowledge about and access to the complete set of possible 
strategies~\cite{NM:1992,NBM:1994,OHLN:2006,TDW:2007,DHD:2014,ALCFMYN:2017}. 
A typically studied game setting is the following: a number of agents 
interact with one another via some kind of network topology (e.g., regular 
or complex), and each agent can take on one strategy from a pre-defined 
strategy set determined according to the typical individual behaviors 
observed from real world systems, leading to classic games such as the 
Prisoner's dilemma games (PDGs)~\cite{HA:1981}, the snowdrift games 
(SGs)~\cite{Sugden:book}, and the public goods games (PGGs)~\cite{Hardin:1968}.
Such a game system typically contains two strategies: cooperation and 
defection, where the latter is a selfish action that usually generates 
immediate higher payoff~\cite{Smith:book}. A remarkable achievement of 
evolutionary game theory is that it resolves this paradox
in a counterintuitive but completely reasonable way. In fact, previous
works have uncovered a variety of cooperation-facilitating mechanisms such as 
reputation and punishment~\cite{FG:2002}, random diffusion~\cite{SFVA:2009}, 
memory effect~\cite{WRCW:2006}, network reciprocity~\cite{NM:1992,HD:2004},
random noise~\cite{SVS:2005,VGS:2006,VGS:2008}, success-driven 
migration~\cite{HY:2009}, asymmetric cost~\cite{DCHW:2009}, teaching 
ability~\cite{SS:2007}, and social or financial 
diversity~\cite{SSP:2008,CHWZW:2009,CL:2012}. In addition to game dynamics with two strategies (i.e., 
cooperation and defection), there are also games with three or more 
strategies, such as the ``rock-paper-scissors'' type of games where any  
strategy has a cyclic advantage over another~\cite{ML:1975,FA:2001,
KRFB:2002,SH:2002,SKM:2003,RMF:2007a,SF:2007,RMF:2008,BRSF:2009,WLG:2010,
SWYL:2010,YWLG:2010,NWLG:2010,NYWLG:2010,WNLG:2011,JWLN:2012,PDHL:2013} 
and extensions~\cite{SS:2004,PA:2008,SSS:2007,SS:2008,ABLMO:2012,VSS:2013,
KPWH:2013,ZYHPDL:2014,PDJL:2017}. Studies of such game dynamics have led
to great insights into species coexistence and biodiversity in complex 
ecosystems. 
Theoretically, methodologies from statistical physics have been used to 
understand complex spatiotemporal game 
dynamics~\cite{ST:1998,SH:2002,HS:2005,SVS:2005,VGS:2008,SB:2016}.

In spite of its widespread use in previous works on evolutionary game 
dynamics, the assumption that every agent (game player) in the system has 
the same strategy set may be idealized. In reality, due to the diversity 
in the knowledge background and personal experience, it is only natural to 
assume heterogeneity in individuals' available strategy sets. It is also 
possible that an individual's strategies are affected by his/her emotions and 
external factors. In addition, individuals playing a game may have quite
different understandings of other's strategies. When the game players do not 
possess an identical strategy set because not everyone has full knowledge
about the complete set of available strategies, the underlying game 
is called hypergame, a term coined by Bennett in 1977~\cite{Bennett:1977}.     
In general, hypergame takes into account the realistic 
situation where players' understandings and choices of the game strategies 
can be different. As a result, hypergame dynamics are capable of modeling 
competitions and conflicts in the real world more closely, leading to better 
and more realistic solutions than the classic game 
dynamics~\cite{BJ:2012,YK:2012,KGL:2015}. 
During a hypergame, an agent has his/her own knowledge base to make judgment 
of the environment and determines which strategy to use.
The true payoff gained by the agent is determined
by his/her current strategy versus the actual strategies in the system. 
In fact, the player's choice of action reflects the way he/she perceives the 
reality and the game outcome, which is usually not accurate. The inaccuracy in 
the perception can affect the evolutionary dynamics of cooperation in a 
fundamental way, and may lead to different phenomena than predicted by
classical evolutionary game dynamics in previous works. 

In this paper, we study hypergame from the perspective of network dynamics. In 
spite of its importance, there has been little previous study of evolutionary 
hypergame dynamics. Because of the system complexity induced by the
uncertainties in individual agent's understanding and choice of the game 
strategies, we seek to construct the simplest possible class of prototypical 
models that retain the essential features of hypergame dynamics to uncover the 
underlying generic behaviors. 
In particular, we consider the system setting where there are three possible 
strategies in the system. To every agent in the system, two of the three 
strategies are available. During the dynamical evolution, at any given time 
an agent can use either one of the two available strategies with certain 
probabilities. The probability of adopting a strategy is thus a key 
(bifurcation) parameter in our model. Our computations reveal that, as the 
bifurcation parameter is changed, our parsimonious model generates 
distinct and more complex dynamical phases than those from the 
classical evolutionary game models. In particular, there are deterministic 
single-strategy-absorption phases, a ``rock-paper-scissors'' type phase, and 
a phase of high uncertainty in which the dominant strategy adopted by the 
population is unpredictable. We obtain an qualitative understanding of 
the emergence of the multiple dynamical phases by exploiting the pair 
approximation and solving the mean-field master equation. 
A striking finding is the phenomenon of strategy revival: the population 
adopting a specific strategy can decrease and approach zero but it can 
revive and dominate at later time. While qualitatively this can be 
explained based on cyclic competitions among three strategies, to develop
a quantitative understanding is an open issue. From the network control perspective, 
our finding that a slight change in the bifurcation parameter can completely 
overturn the relative advantage between the strategies suggests that the 
complex game dynamics can be harnessed through small perturbations to the
parameter.

\section{Evolutionary hypergame model} \label{sec:model}

To construct a parsimonious model of evolutionary hypergame
dynamics, we begin with the generalized prisoner's dilemma game (gPDG) with 
three strategies: cooperation ($C$), defection ($D$), and loneliness ($L$). 
If an agent adopts the $L$ strategy, he/she does not actually participate 
in the game but nonetheless is guaranteed to receive a low payoff. 
For simplicity, we use the prisoner's dilemma game model 
introduced by Nowak and May~\cite{NM:1993}, which captures the essential 
feature of hypergame. The payoff matrix $M$ is given by
\begin{equation} \label{eq:PM}
\left(\begin{array}{cccc}
    & C & D & L\\
    C & 1 & 0 & 0\\
    D & b & 0 & 0\\
    L & \sigma & \sigma & \sigma\\
\end{array}\right)
\end{equation}

Because agents have a different understanding of the competition environment, 
during the hypergame, each agent is able to distinguish and adopt only two 
of the three strategies. For each agent, the resource is constrained, so the 
two strategies have weights that sum up to unity. There are thus three
types of agents in the model: agents having available strategies (1) $C$ and
$D$, (2) $D$ and $L$, and (3) $L$ and $C$, respectively. For each of the three combinations, the probability of adopting the first strategy is $\rho$   
while that adopting the second strategy is $1-\rho$. 
For each agent, the strategy set is thus restrictively mixed because there 
is a missing strategy. For the whole system, there are then three distinct 
such strategies. Mathematically, the three strategies can be represented 
by the following three vectors:  
\begin{eqnarray} 
\nonumber
S^{(1)} & = & \left(\begin{array}{cc}
        \rho \\
        1-\rho\\
        0\\
        \end{array}\right), \ \ \ 
S^{(2)}=\left(\begin{array}{cc}
        0  \\
        \rho \\
        1-\rho\\
        \end{array}\right), \\ \label{eq:vectors}
S^{(3)} & = &\left(\begin{array}{cc}
        1-\rho \\
        0\\
        \rho\\
        \end{array}\right),
\end{eqnarray} 
where rows 1, 2, and 3 indicate the adoption probabilities of strategies 
$C$, $D$, and $L$, respectively.

In the simulations, agents are placed on a square lattice with periodic 
boundary conditions. At each time step, each agent plays gPDG with its 
nearest neighbors. The total payoff gained is the sum of the payoffs 
from playing the game with all its neighbors, which is given by
\begin{equation} \label{eq:payoff}
U_n=\sum\limits_{m}u_{nm}=\sum\limits_{n,m}{S_n}^{T}MS_m,
\end{equation}
where $U_n$ denotes the total payoff of agent at lattice node $n$, $u_{nm}$ 
is the payoff obtained by agent $n$ while playing the game with agent at  
lattice node $m$, $M$ is the payoff matrix in Eq.~(\ref{eq:PM}), $S_n$ and 
$S_m$ are the strategy vectors of the two agents, respectively. After 
obtaining the payoff, agent $n$ with strategy $S_n$ is replaced by agent 
$m$ with the probability given by the Fermi rule~\cite{CHWZW:2009}:
\begin{equation}
P_{S_n\longrightarrow S_m}=\frac{1}{1+\exp[(U_m-U_n)/\kappa]},
\end{equation}
where $\kappa$ measures the stochastic uncertainties (noise) characterizing 
irrational choices.

In each dynamical realization, initially $N$ agents with the three types 
of strategies are randomly distributed in the square lattice with equal 
probability, i.e., $F_1 = F_2 = F_3 = 1/3$, where $F_x$ is defined in 
Eq.~(\ref{DFF}). 
The system evolves in time until an equilibrium is reached.
To be concrete, we set the game parameters as $b=1.02$, $\kappa=0.1$, and 
$\sigma=0.25$. We check to ensure that reasonably different choices of the 
parameters lead to qualitatively identical behaviors from the simulations.
The adoption probability $\rho$ is a key parameter in the system, which 
is chosen as the control or bifurcation parameter.

\begin{figure}
\centering
\includegraphics[width=\linewidth]{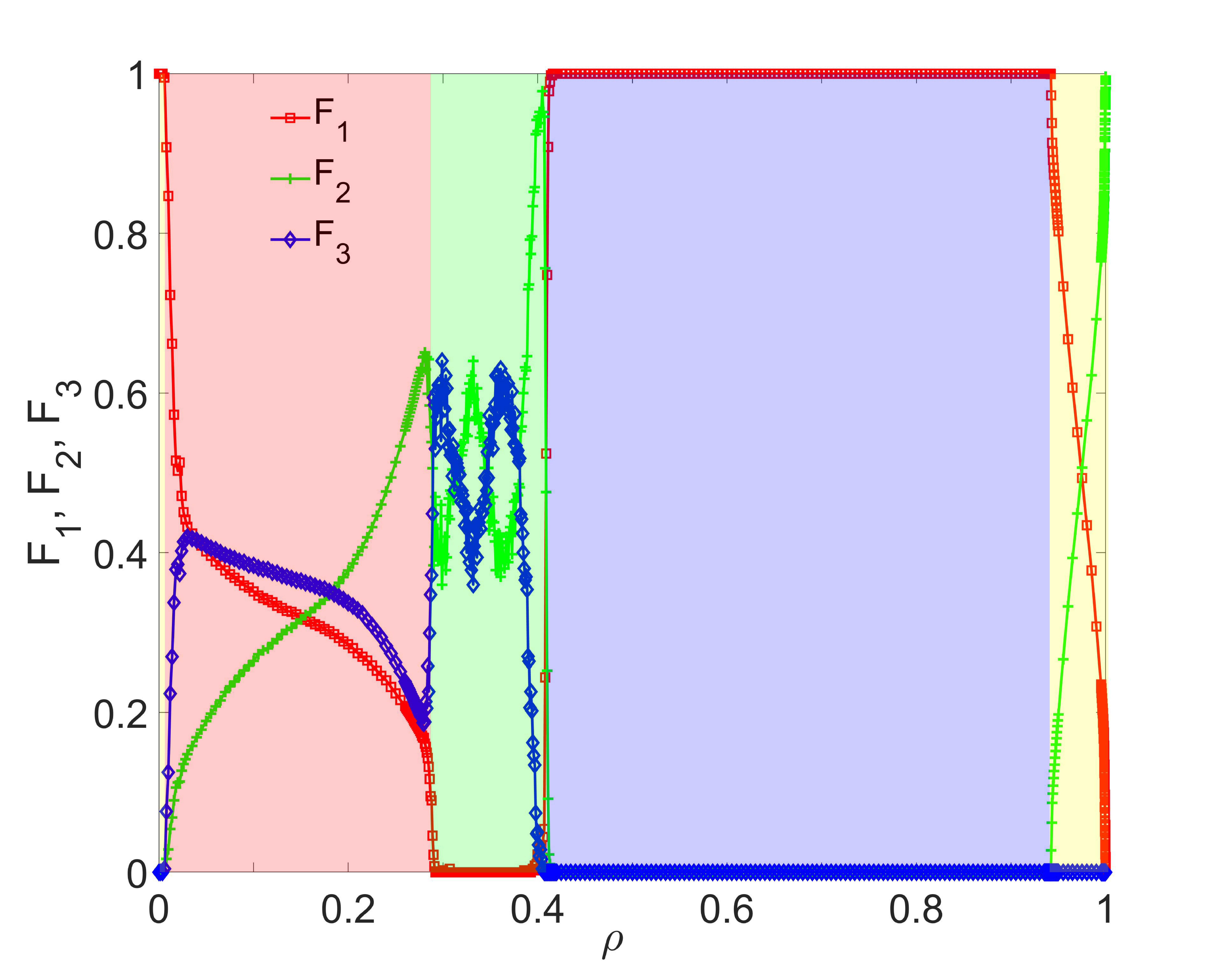}
\caption{ (Color online) {\em Dependence of agent frequency on the bifurcation 
parameter}. The system is a $70 \times 70$ square lattice with periodic 
boundaries. The number of agents is $N=4900$. The red (medium gray), green 
(light gray) and blue (dark gray) curves show the frequencies $F_1$, $F_2$, 
and $F_3$ of the three agents versus the bifurcation parameter $\rho$, 
respectively. For different values of $\rho$, the system evolves into a final 
equilibrium state in different time $T$. In particular, for 
$0\leq\rho\leq0.03$, $0.26\leq\rho\leq0.3$, $0.935\leq\rho\leq0.95$, and 
$0.995\leq\rho\leq1$, we have $T = 5 \times 10 ^5$. For $0.3<\rho<0.42$, the 
system approaches a final state for $T < 10^4$. For other values of $\rho$, 
the system needs $T = 10^4$ time steps to approach the final state. The 
values of $F_1$, $F_2$, and $F_3$ shown are the results of averaging over 
100 ensembles (random realizations) for most values of $\rho$ except for 
$0.3 < \rho < 0.42$, where 500 ensembles are used because of the relatively 
strong statistical fluctuations in this parameter interval. Depending on the 
behavior of the equilibrium strategy frequencies, the whole parameter interval 
can be divided into four regions: (1) $0 < \rho \alt 0.03$ and 
$0.95 \agt \rho < 1$ (region 1), (2) $0.03 \alt \rho \alt 0.3$ (region 2), 
(3) $0.3 \alt \rho \alt 0.4$ (region 3), and (4) $0.4 \alt \rho \alt 0.95$ 
(region 4)}.
\label{fig:allrho}
\end{figure}

\section{Numerical results} \label{sec:numerics}

\subsection{The equilibrium state}

The equilibrium frequency of the restrictively mixed strategy 
$S^{(x)}$ ($x=1,2,3$) on a 
square lattice of $N$ nodes is given by
\begin{equation}\label{DFF}
F_x = \frac{\sum\limits_{n=1}^{N}I(S_n,S^{(x)})}{N},
\end{equation}
{where $S_n$ denotes the strategy of agent at lattice node $n$} and 
$I(S_n,S^{(x)})$ is an indicator function [$I(S_n,S^{(x)}) = 1$ for 
$S_n=S^{(x)}$ and $I(S_n,S^{(x)}) = 0$ otherwise]. 
We use 100 random simulation realizations to obtain the value of $F_x$.
Figure~\ref{fig:allrho} shows the frequencies $F_1$, $F_2$, and $F_3$ 
associated with the strategies $S^{(1)}$, $S^{(2)}$, and $S^{(3)}$, 
respectively, versus the parameter $\rho$. In different regions of $\rho$
value, the frequencies exhibit dramatically different behaviors. Specifically,
for $\rho \agt 0$, $S^{(1)}$ dominates the entire population, while the 
frequencies of $S^{(2)}$ and $S^{(3)}$ are essentially zero. As $\rho$ 
is increased, a sharp reduction in $F_1$ occurs, leading to an increase in
the values of both $F_2$ and $F_3$. At this point, the system enters into 
a state in which strategies coexist with similar frequency values. When 
$\rho$ reaches the value of about $0.3$, the value of $F_1$ becomes effectively
 zero, while those of $F_2$ and $F_3$ alternate. For $\rho\approx 0.4$,
$F_2$ peaks, indicating that $S^{(2)}$ now dominates the system. With a small 
increment in the value of $\rho$, $S^{(1)}$ becomes dominant, after which 
the system remains in the $S^{(1)}$-absorption state for a large range of
the $\rho$ value until about $\rho \approx 0.94$ when the gradual increase 
and decrease in the values of $F_2$ and $F_1$, respectively, drive the 
system into an $S^{(2)}$-absorption state at $\rho=1$. We thus see that, as 
$\rho$ is changed, the system shows highly complex behaviors with distinct 
evolutionary patterns. For example, one strategy may have superior advantage 
over the others in some parameter region but may lose appeals completely 
in other regions. The phase transitions associated with most of the dramatic
changes in the system dynamics are abrupt. This numerical finding indicates 
that hypergame dynamics can be much richer than conventional game dynamics 
with pure strategies.

For convenience, we use different colors to denote the four regions with 
qualitatively different behaviors (i.e., different phases), as shown 
in Fig.~\ref{fig:allrho}. From the model setting, we
see that each of the three strategy vectors can get infinitesimally close 
to another for $\rho\rightarrow 0$ or $\rho\rightarrow 1$: 
$S^{(1)}|_{\rho\rightarrow0} \rightarrow S^{(2)}|_{\rho\rightarrow1}$, 
$S^{(2)}|_{\rho\rightarrow0} \rightarrow S^{(3)}|_{\rho\rightarrow1}$, 
and $S^{(3)}|_{\rho\rightarrow0} \rightarrow S^{(1)}|_{\rho\rightarrow1}$,
providing an explanation for the similarities in the behaviors of $F_1$, $F_2$, 
and $F_3$ for $\rho\rightarrow 0$ and $\rho\rightarrow 1$. For this reason,
the regions corresponding to $\rho\rightarrow0$ and $\rho\rightarrow1$ are 
marked with the same color. 

\begin{figure}
\centering
\includegraphics[width=1\linewidth]{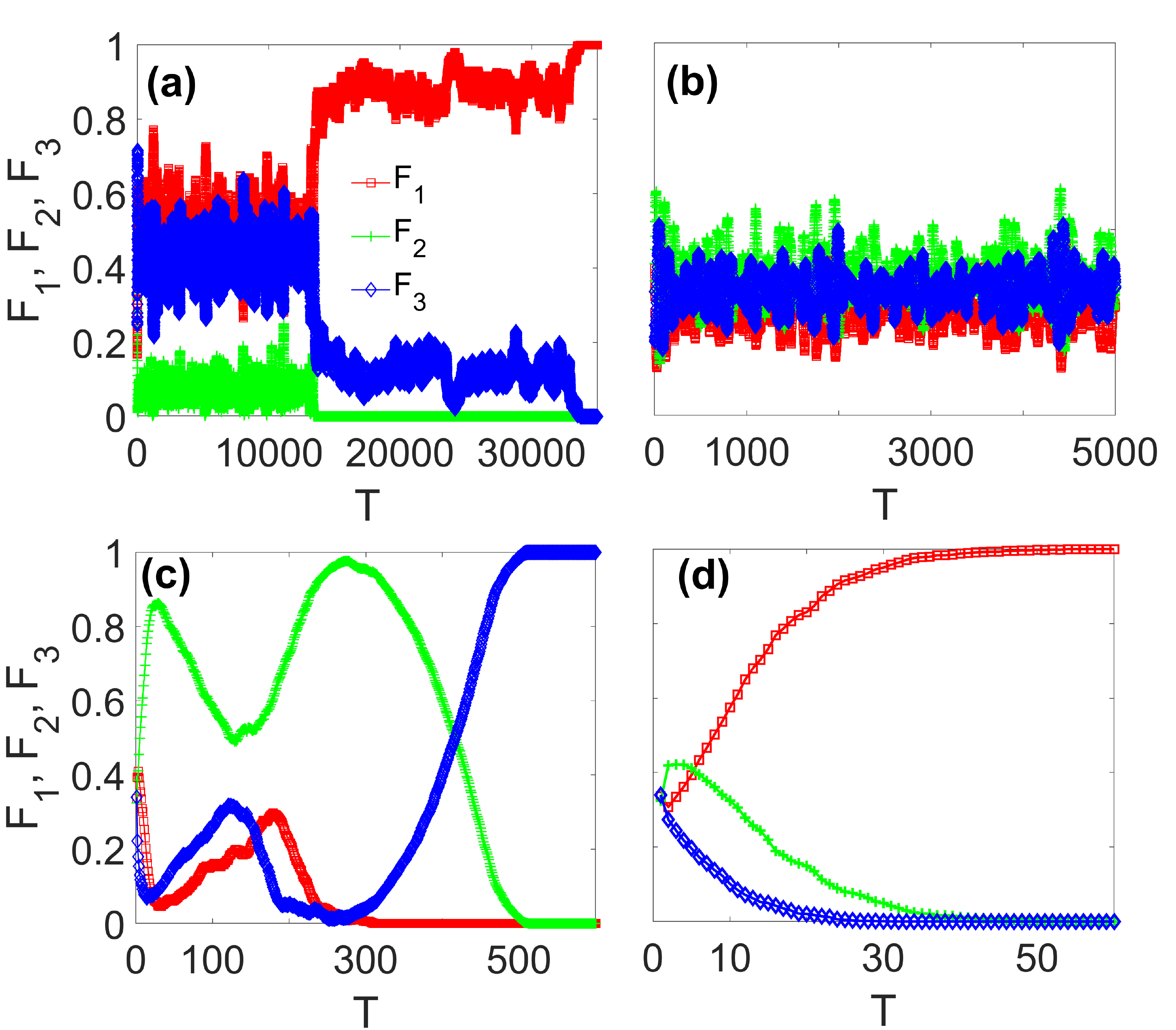}
\caption{ (Color online) {\em Transient behaviors associated with hypergame 
dynamics}. (a-d) Transient behaviors in the evolution of the restrictively 
mixed strategies for $\rho=0.001$ (in region 1), $\rho=0.2$ (in region 2), 
$\rho=0.31$ (in region 3) and $\rho=0.7$ (in region 4), respectively. The 
red (medium gray), green (light gray), and blue (dark gray) curves represent 
the fractions of $S^1$, $S^2$, and $S^3$, respectively, where $T$ is the 
number of evolution time steps. The results from all four panels are from 
a single realization of the system. The system parameters are the same as 
those in Fig.~\ref{fig:allrho}.}
\label{fig:transient}
\end{figure}

\subsection{Transient behaviors}

The transient behaviors that the system exhibits before approaching the 
equilibrium reveal more about the hypergame dynamics than the equilibrium
itself. For example, in region 1 ($0 < \rho \alt 0.03$ and 
$0.95 \agt \rho < 1$), there are two restrictively mixed strategies: 
$S^{(1)}$ and $S^{(3)}$ on the left side of $\rho=0$ (or $S^{(2)}$ and 
$S^{(3)}$ on the right side of $\rho=1$), where the strategy with a stronger 
defective weight gains evolutionary advantage over the one with more 
cooperative weight while the third strategy disappears long before the 
equilibrium is reached, as shown in Fig.~\ref{fig:transient}(a). 
This is due to that, when the value of $\rho$ is near zero or unity, the 
restrictively mixed strategies are similar to the pure strategies in 
traditional game dynamics, where the defective strategy dominate on networks 
with a homogeneous topology due to fluctuations and the finite size effect. 
In this case, the third strategy has little chance to lead to high payoff and 
would be eliminated quickly, so the final state is the coexistence of 
two restrictively mixed strategies. In region 2 ($0.03 \alt \rho \alt 0.3$), 
the final state is the coexistence of all three restrictively mixed 
strategies.
In this case, no strategy can be eliminated, as shown in 
Fig.~\ref{fig:transient}(b). The 
rapid increase in the frequency of $S^{(2)}$ for $\rho$ close to $0.1$ 
indicates that a slight increment in $\rho$ can reactivate the strategy 
$S^{(2)}$ and a small amount of defection can lead to a substantial increase 
in the evolutionary fitness. As the value of $\rho$ approaches $0.3$, the 
strategy $S^{(2)}$ becomes relatively dominant, while the strategies
$S^{(1)}$ and $S^{(3)}$ approach extinction, as shown in 
Fig.~\ref{fig:transient}(c). As $\rho$ passes through a certain threshold
value, the system enters into region 3 ($0.3 \alt \rho \alt 0.4$), in which 
the transient behavior can be complicated but the equilibrium falls into only
one of two states: the $S^{(2)}$ or the $S^{(3)}$ absorption state. This 
means that a single strategy always wins, either $S^{(2)}$ or $S^{(3)}$. The 
frequencies of $S^{(2)}$ and $S^{(3)}$ shown in Fig.~\ref{fig:allrho} 
are ensemble averaged values of the number of agents in the $S^{(2)}$ 
and $S^{(3)}$ absorption states, respectively, where a single realization 
leads to a state with only one strategy: $S^{(2)}$ or $S^{(3)}$. Strikingly, 
before the equilibrium is reached, $S^{(1)}$ and $S^{(3)}$ enter into a 
nearly extinction state, where their frequencies are so low that random 
fluctuations can eliminate one of them. However, if $S^{(1)}$ becomes extinct, 
$S^{(3)}$ will take over the entire system. Figure~\ref{fig:transient}(c) 
also shows the dramatic change in the frequency of $S^{(3)}$. The 
coexistence of the three strategies lasts longer for a larger network,
so $S^{(1)}$ and $S^{(3)}$ are more resilient to random fluctuations.
In region 4 ($0.4 \alt \rho \alt 0.95$), $S^{(1)}$ take over after it wins 
the competition with $S^{(2)}$, while $S^{(3)}$ survives only in the first 
few time steps. The final state is one dominated by $S^{(1)}$.

\begin{figure}
\centering
\includegraphics[width=1\linewidth]{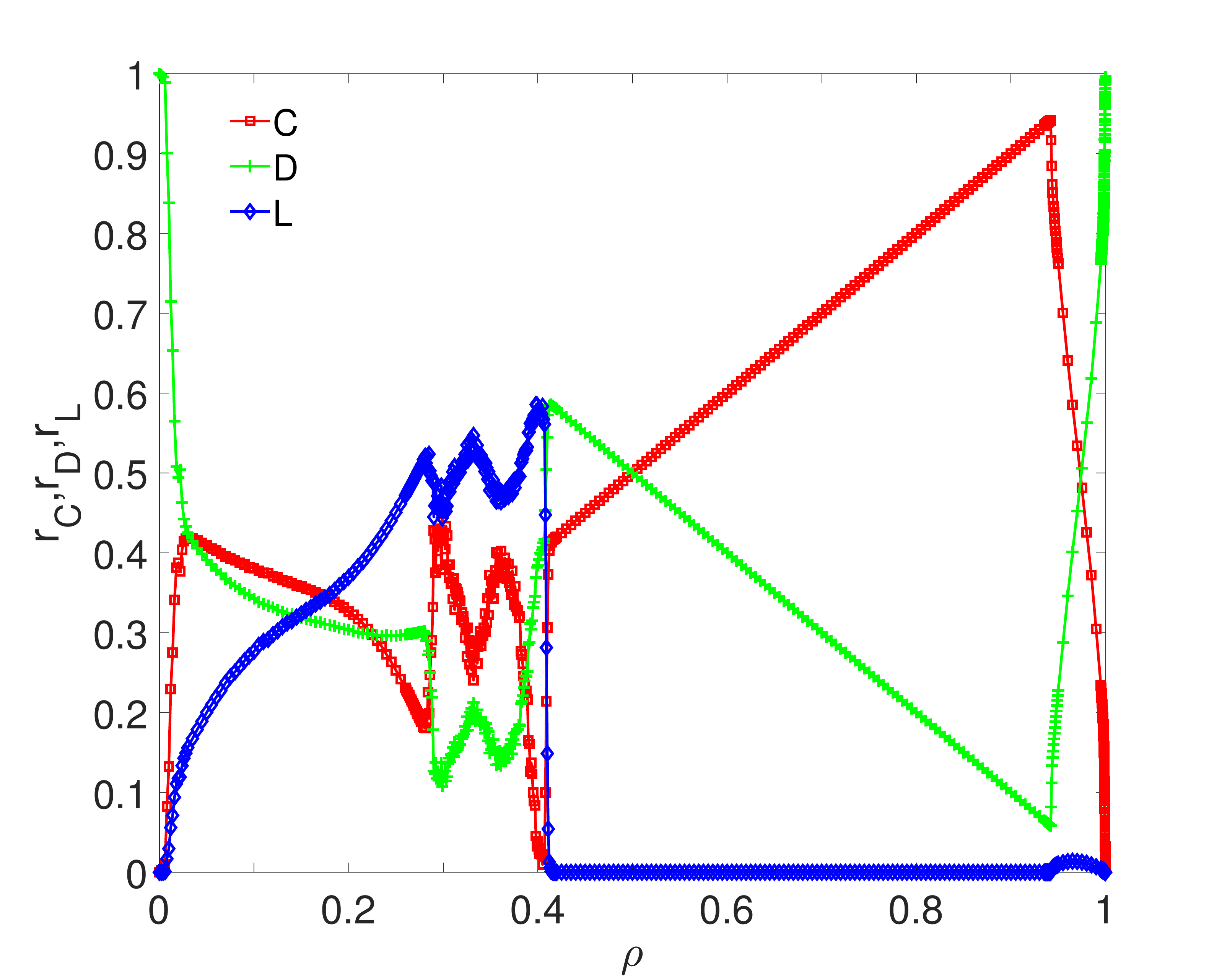}
\caption{ (Color online) {\em Actual fractions of cooperation, defection, and 
loneliness}. These fractions are denoted by $r_C$, $r_D$, $r_L$, respectively, 
which depend on the value of $\rho$ in a somewhat complicated manner. However, 
there are parameter regions in which cooperation dominates, e.g., 
$0.5 \alt \rho \alt 0.95$. Simulation parameters are the
same as those in Fig.~\ref{fig:allrho}.}
\label{fig:Actual_CDL}
\end{figure}

\subsection{Fraction of cooperation}

The fraction of cooperation is a more intuitive characterizing quantity of
the system dynamics. Figure~3 shows the actual fractions of cooperation, 
defection, and loneliness versus $\rho$. While the dependence of the fraction
of cooperation on $\rho$ is complicated, there are parameter regions in
which cooperation dominates, e.g., $0.5 \alt \rho \alt 0.95$,
For $\rho \approx 0.95$, the fraction of cooperation is nearly one.

\begin{figure}
\centering
\includegraphics[width=1\linewidth]{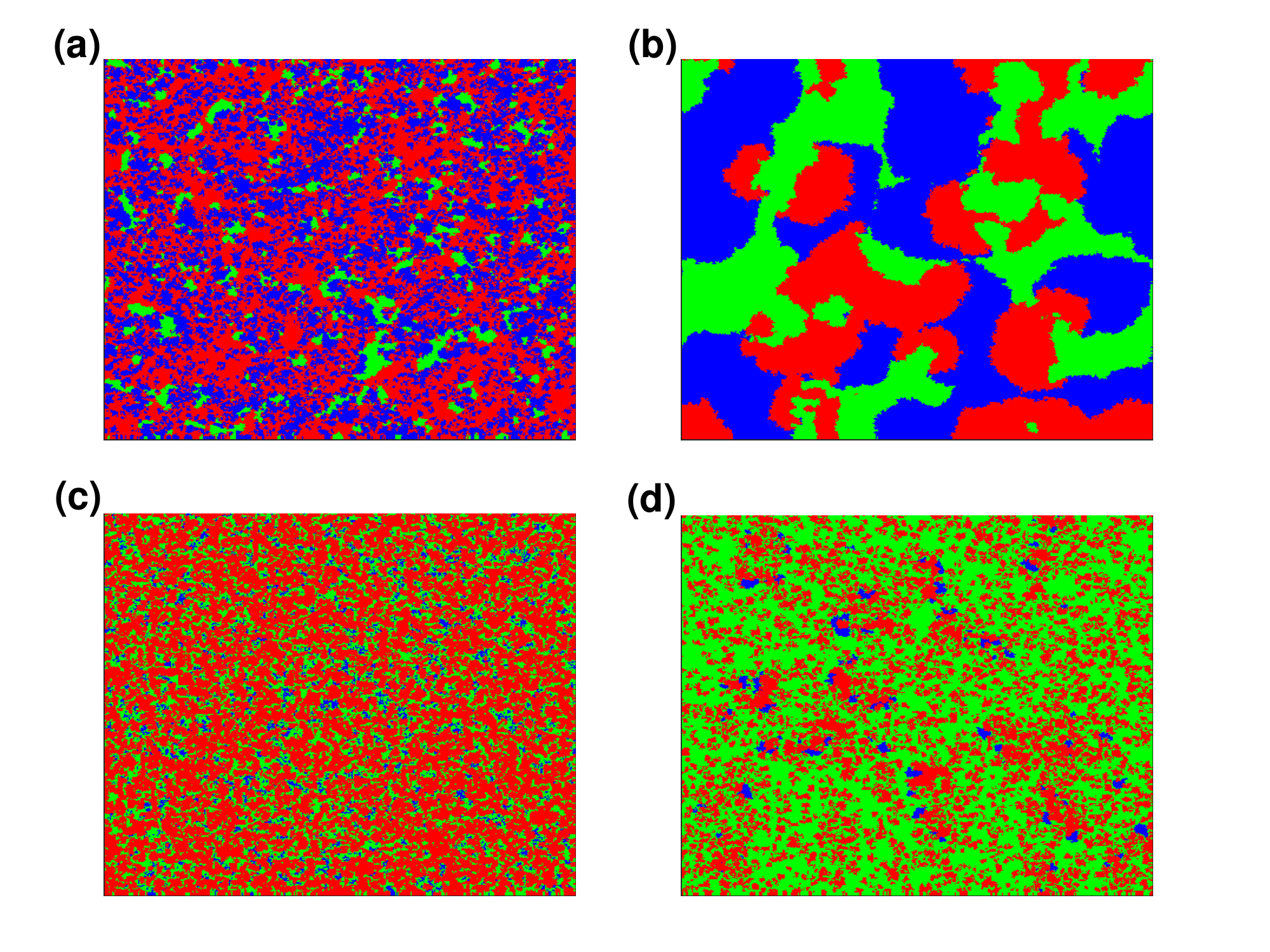}
\caption{ (Color online) {\em Snapshots of typical lattice configuration for 
different $\rho$ values}. 
(a-d) Snapshots of self-organizing patterns on the lattice associated
with equilibrium coexistence of $S^{(1)}$ [red (medium gray)], $S^{(2)}$ 
[green (light gray)], and $S^{(3)}$ [blue (dark gray)] for 
$\rho=0.01, 0.35, 0.65$ and $0.99$, respectively. Other parameters are 
$b=1.02$, $\sigma=0.25$, and $K=0.1$. The lattice size is $400\times 400$.}
\label{fig:Pattern1}
\end{figure}

\begin{figure*}
\centering
\includegraphics[width=\linewidth]{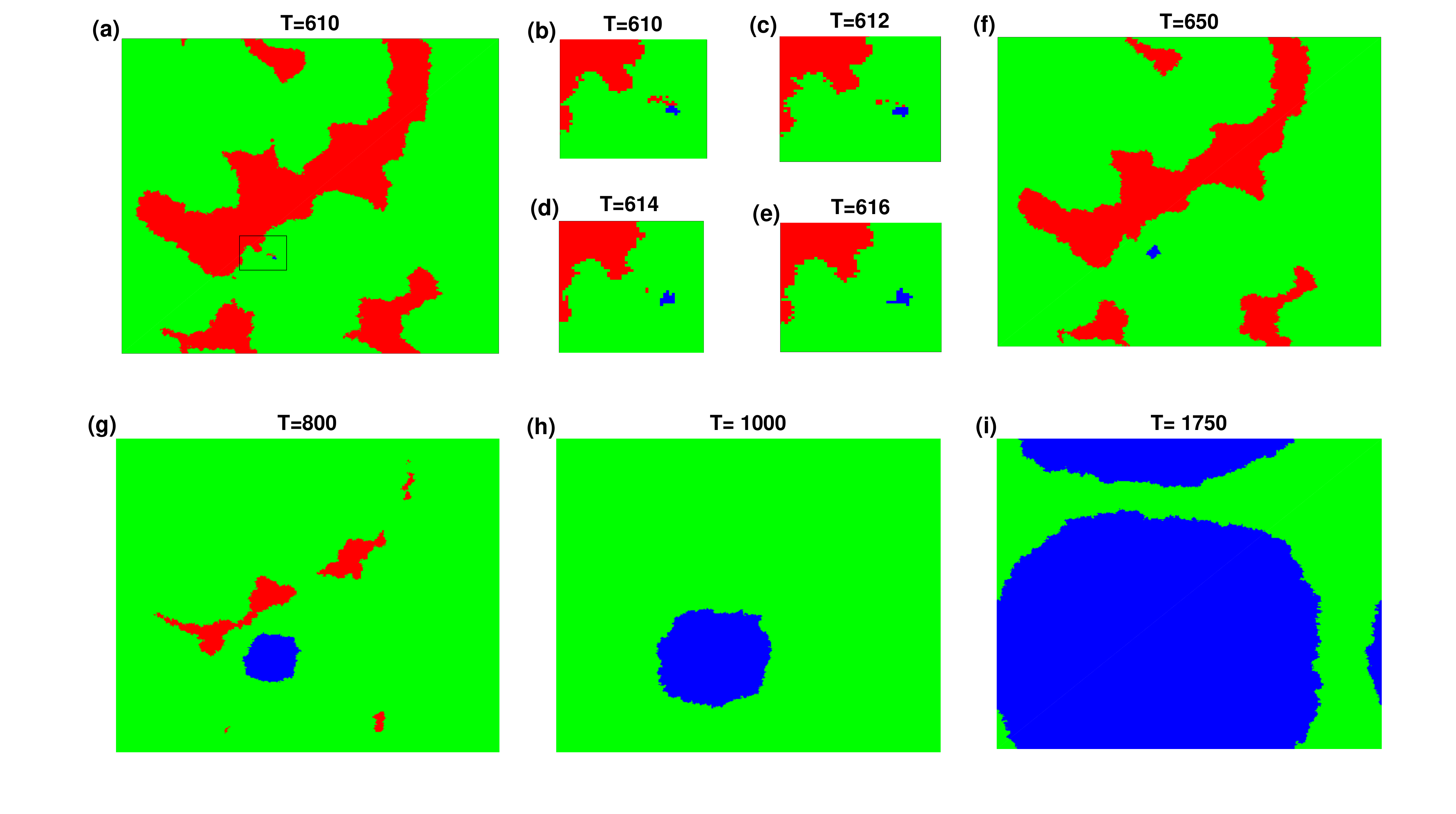}
\caption{ (Color online) {\em Strategy revival}. 
(a,f,g,h,i) For $\rho=0.35$, the patterns of $S^{(1)}$ [red (medium gray)], 
$S^{(2)}$ [green (light gray)], $S^{(3)}$ [blue (dark gray)] on the square 
lattice of size $400\times400$. (b-e) Magnification of the patterns at 
different time. Other parameters are the same as in Fig.~\ref{fig:Pattern1}.}
\label{fig:Dominatepattern}
\end{figure*}

\begin{figure}
\centering
\includegraphics[width=\linewidth]{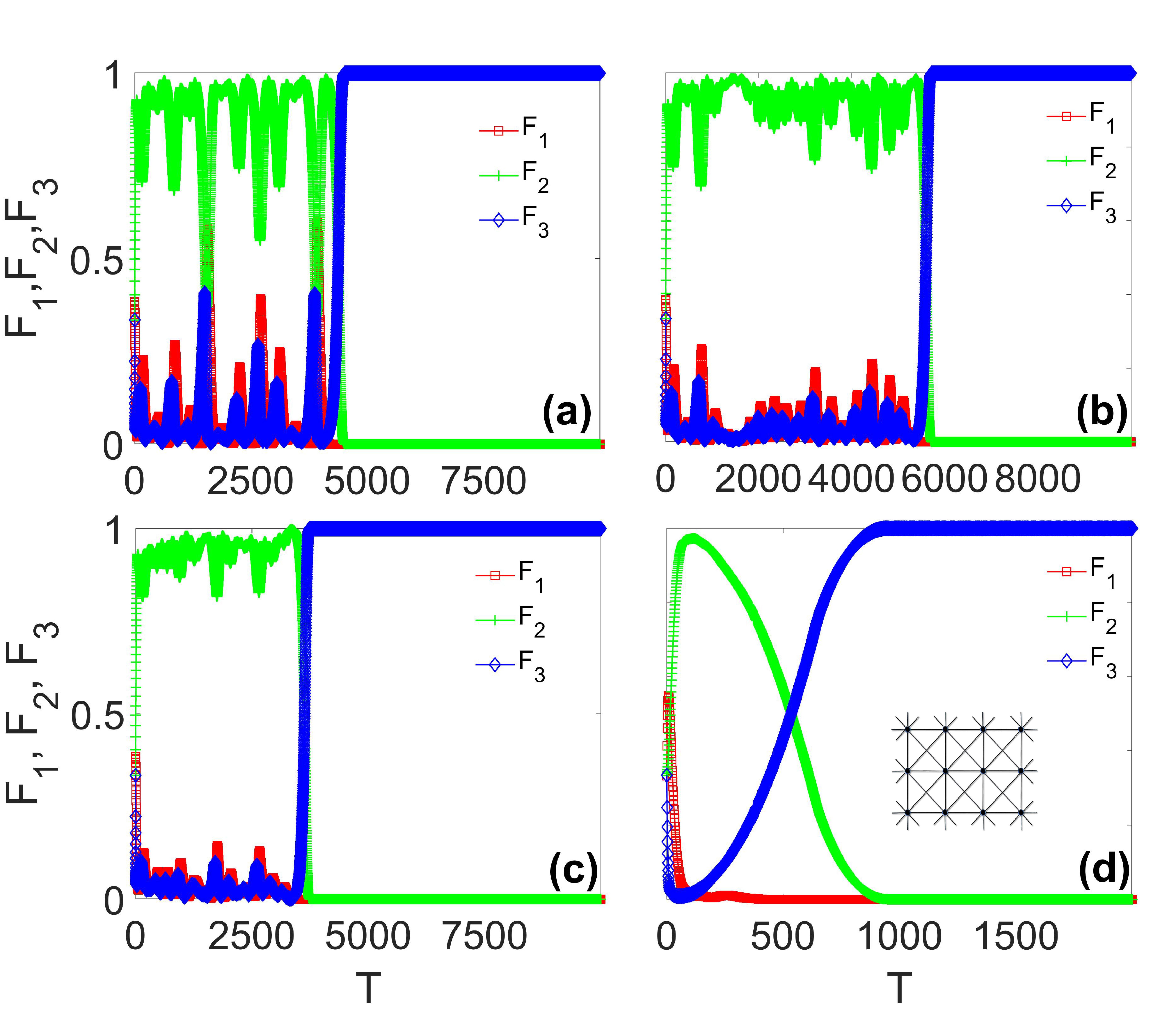}
\caption{  (Color online) {\em Persistence of strategy revival in small world 
networks}. (a-c) The small world networks are generated by randomly rewiring 
$1\%$, $1.5\%$, $2\%$ of the links of the square lattice, respectively. In all 
cases, there is strategy revival, in spite of an increase in the time that 
it takes for it to occur. This time can be reduced by increasing the average
degree of the network, as shown in (d) for $\langle k\rangle = 8$. The network
size for (a-d) is $160000$. The value of the parameter $\rho$ is 0.35 
for (a-c) and 0.325 for (d). All other parameters are the same as in 
Fig.~\ref{fig:Dominatepattern}.}      
\label{fig:Diff_Lattice_DP}
\end{figure}

\begin{figure}
\centering
\includegraphics[width=\linewidth]{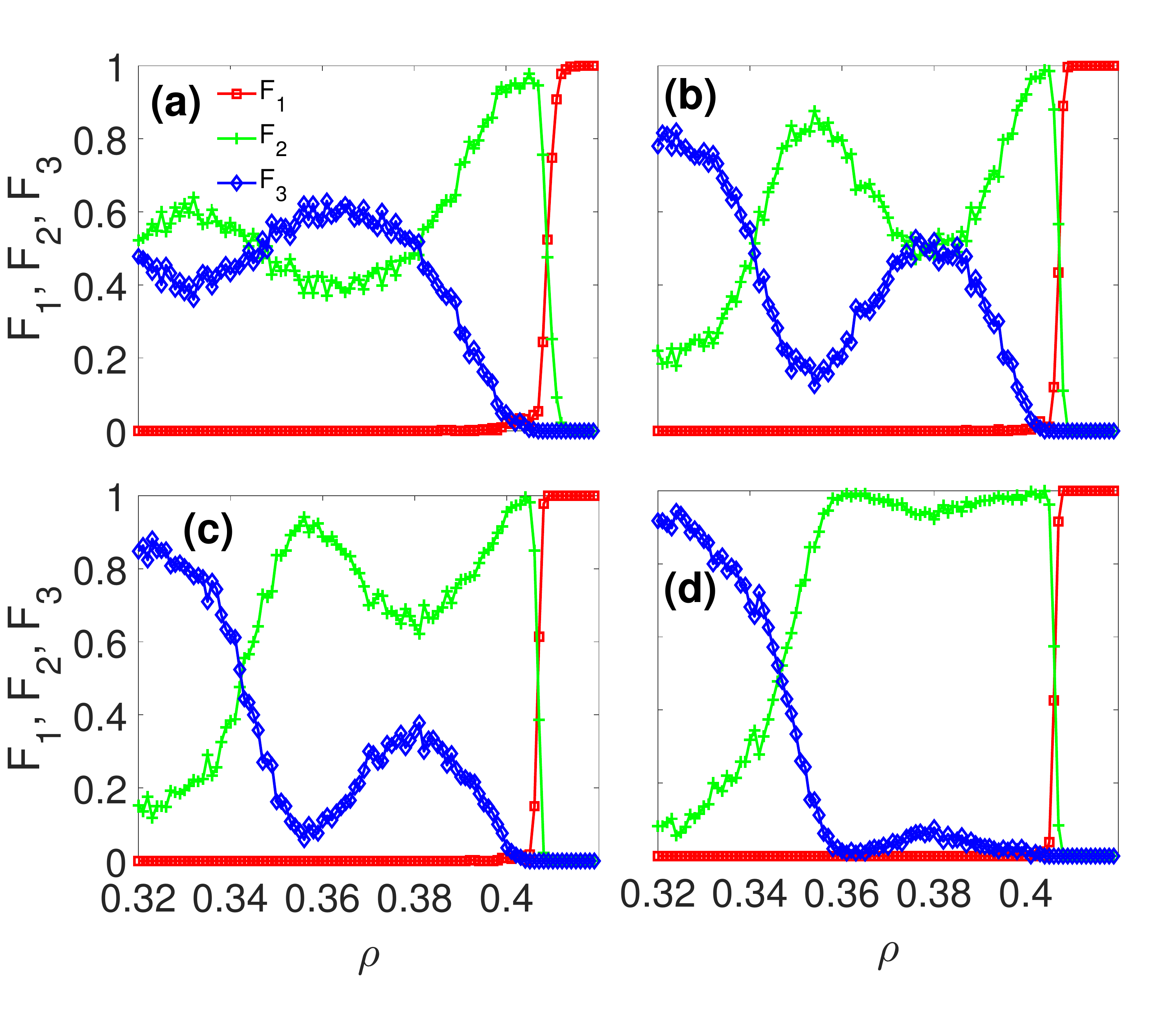}
\caption{ (Color online) {\em Frequency of a dominant state.} 
The frequency values for the system to exhibit an $S2$ or an $S3$ dominant 
state. All data points are the result of averaging over 500 statistical 
ensembles. The red (medium gray), green (light gray), and blue (dark gray) 
curves represent $S1$, $S2$, and $S3$, respectively. (a-d) The frequencies 
of $S2$ and $S3$ for a $70\times 70$, $200\times200$, $250\times250$, and 
$400\times400$ lattice, respectively. Periodic boundary conditions are used.} 
\label{fig:random_dominate}
\end{figure}

\subsection{Evolution pattern on lattice and strategy revival}

To further understand the coexistence of strategies for values of $\rho$ in 
distinct dynamical regimes, we compute the patterns of the equilibrium 
coexistence states on the lattice, as shown in Fig.~\ref{fig:Pattern1}. 
In region 1, on the right side of $\rho=0$, before an equilibrium state is 
reached, a typical pattern is that $S^{(3)}$ forms small but relatively 
stable clusters with irregular boundaries distributed evenly on the lattice. 
In between the clusters is $S^{(1)}$, as shown in Fig.~\ref{fig:Pattern1}(a). 
A similar phenomenon occurs on the near left of $\rho=1$ for strategies 
$S^{(2)}$ and $S^{(1)}$. In region 2, $S^{(1)}$ and $S^{(3)}$ form large 
clusters with regular boundaries, while $S^{(2)}$ acts as the background of 
those clusters, as shown in Fig.~\ref{fig:Pattern1}(b). Frequent strategy 
transitions occur on the boundaries. For example, on the boundary between 
$S^{(1)}$ and $S^{(2)}$, the probability of $S^{(1)}$ transforming 
into $S^{(2)}$ is higher than that of the transformation in the opposite
direction. On the boundary between $S^{(2)}$ and $S^{(3)}$, $S^{(3)}$ 
is more likely to replace $S^{(2)}$, and on the boundary between $S^{(1)}$ 
and $S^{(3)}$, $S^{(1)}$ is more likely to exclude $S^{(3)}$. This interaction 
pattern is effectively that of a cyclic (rock-paper-scissors - RPS) game. 
Closer to region 3, the frequencies of $S^{(1)}$ and $S^{(3)}$ in the 
equilibrium states decrease continuously, indicating the possible occurrence 
of a region in which only $S^{(2)}$ exists as $\rho$ is increased. However, 
this cannot occur for the reason that, in region 3, the existence of $S^{(3)}$ 
is robust, as shown in Fig.~\ref{fig:Pattern1}(b). In particular, in this 
parameter region, for a long period of time, the transient evolution patterns 
of the three strategies are similar to those associated with the RPS-like 
game in which all three strategies coexist. Surprisingly, after a long 
time evolution, $S^{(1)}$ or $S^{(3)}$ can suddenly disappear as their 
frequencies approach zero, which breaks the symmetry: if $S^{(1)}$ 
becomes extinct first, $S^{(3)}$ would eventually take over the entire 
network since $S^{(3)}$ is more likely to replace $S^{(2)}$. Likewise,
if $S^{(3)}$ disappears before $S^{(1)}$, the $S^{(2)}$ absorption state 
will finally be realized, since there is a higher probability for $S^{(2)}$
to exclude $S^{(1)}$. The emergence of distinct equilibrium states, namely
the $S^{(2)}$ or the $S^{(3)}$ absorption state, depends on which strategy
[$S^{(1)}$ or $S^{(3)}$] becomes extinct first. In region 4, there is 
no clustering behavior, and $S^{(1)}$ takes over the entire lattice 
rapidly. (See Supplementary Videos~\cite{Supple_Video} for a vivid 
presentation of the different evolution processes.)

Figure~\ref{fig:Dominatepattern} shows a concrete example of the 
striking phenomenon of strategy revival: 
$S^{(3)}$ takes over the entire lattice system 
even when it has become almost extinct at a time (in region 3). 
In particular, Fig.~\ref{fig:Dominatepattern}(a) shows that the system 
can reach a state in which there is only a single $S^{(3)}$ cluster of 
extremely small size in contact with a small size $S^{(1)}$ cluster. 
Figures~\ref{fig:Dominatepattern}(d-f) show the evolution pattern 
after the state in Fig.~\ref{fig:Dominatepattern}(c) has been reached. 
We see that, the smaller $S^{(1)}$ cluster first collapses into several 
components and the part still in contact with $S^{(3)}$ disappears, 
leaving a cluster of $S^{(3)}$ surrounded by $S^{(2)}$ only, while the 
other $S^{(1)}$ regions are surrounded by $S^{(2)}$. The $S^{(1)}$ and 
$S^{(3)}$ clusters become well separated, leading to the dominance of 
$S^{(3)}$: $S^{(1)}$ would eventually be replaced by the surrounding 
$S^{(2)}$. When no $S^{(1)}$ is left, no matter how few $S^{(3)}$ 
holders there are, they will exclude all the $S^{(2)}$ holders and 
overturn the whole square lattice into the $S^{(3)}$ absorption state.
Without the separation, $S^{(3)}$ would be completely excluded by 
$S^{(1)}$ at certain time. In this case, when there are only $S^{(1)}$ 
and $S^{(2)}$ left, $S^{(2)}$ will take over by excluding all $S^{(1)}$. 
This is the mechanism by which the $S^{(2)}$ absorption state is generated.

Figure~\ref{fig:Diff_Lattice_DP} shows that the phenomenon of strategy revival 
can also occur in networks with a more complex structure than a regular 
lattice. In particular, the networks in Figs.~\ref{fig:Diff_Lattice_DP}(a-c) 
are constructed from a regular square lattice with three different percentages 
of link rewiring to generate long range random links - they are small world 
networks. There is strategy revival in all three networks, in spite of the 
increase in the time for the system to reach the state in which the strategy 
$S^{(1)}$ is extinct and eventually replaced by the strategy $S^{(3)}$. The 
time can be reduced by increasing the average degree of the network, as 
exemplified in Fig.~\ref{fig:Diff_Lattice_DP}(d).

To better visualize the spatiotemporal evolution of patterns, we provide
four Supplementary movies~\cite{Supple_Video}. 

\subsection{Emergence of a dominant state}

Figure~\ref{fig:random_dominate} shows the frequencies of the three
restrictively mixed strategies 
versus the control parameter $\rho$. There exist
parameter regions where one of the strategies dominates. For 
$0.32 \alt \rho \alt 0.4$, the dominant strategy can be either $S^{(2)}$ or $S^{(3)}$.
The frequencies also depend on the system size. For small systems, the 
frequency values for $S^{(2)}$ and $S^{(3)}$ (green and blue curves) oscillate about 
the value of $0.5$ as $\rho$ is varied in the interval. There are three 
distinct points of $\rho$ at which the two probabilities are equal.
As the system size is increased, there exists only one such value of 
$\rho$. For example, for $\rho\in[0.32,0.36]$, the probability for $S^{(3)}$
to be dominant increases with system size. For $\rho\in[0.36,0.4]$, $S^{(2)}$ 
will dominate for relatively large systems. 

\section{Theory} \label{sec:theory}

\begin{figure}
\centering
\includegraphics[width=1\linewidth]{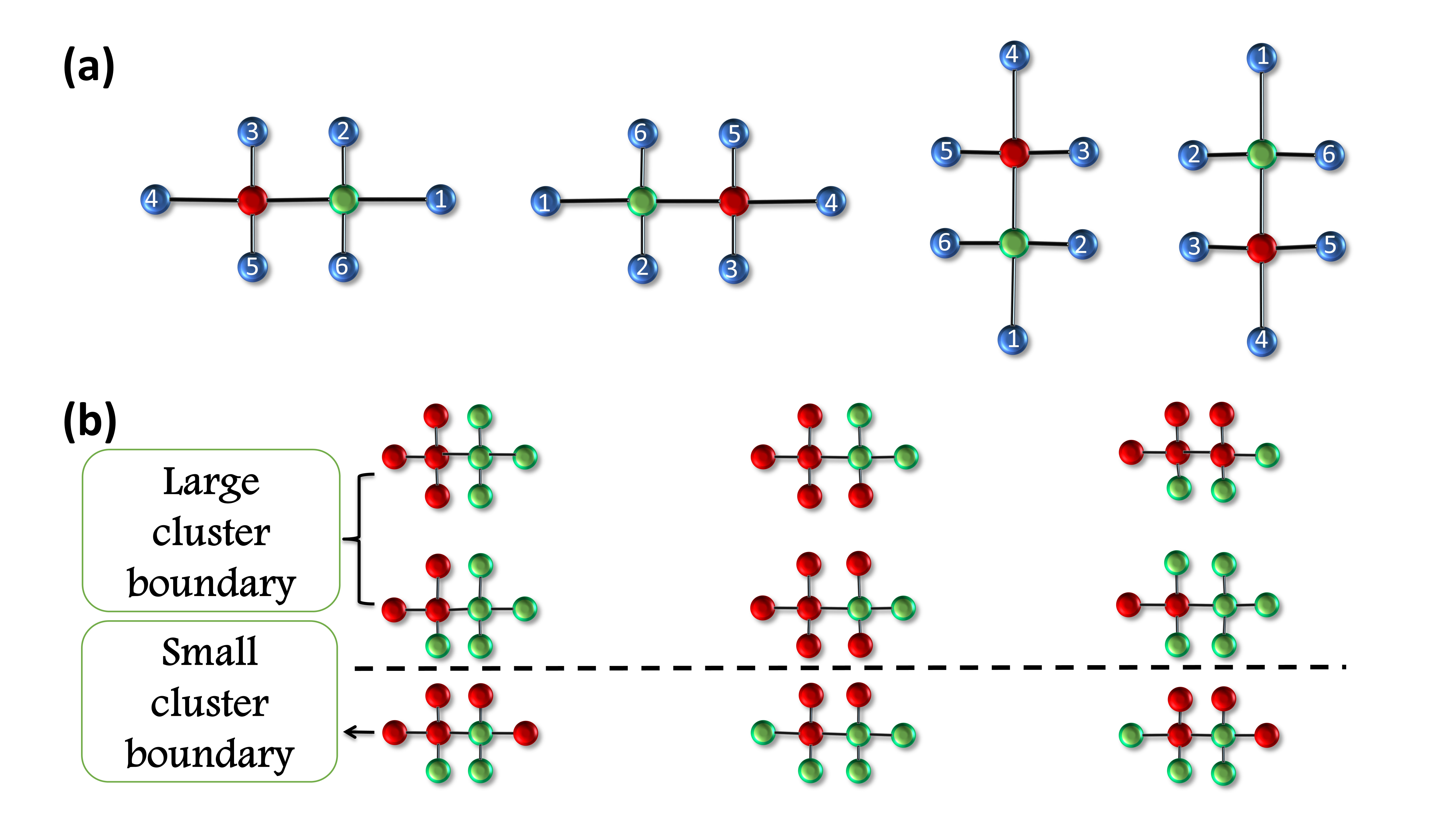}
\caption{ (Color online) {\em Boundary structure of clusters and agents in an 
eight-node motif}. In (a), the red (dark gray), green (light gray), and 
blue (medium gray) nodes represent the 
target agent, the opponent node, and the neighbors of this pair of nodes,
respectively. (b) The cluster boundary structure, where there are 
two types of agents. The upper part of the panel appears on the boundary 
of the large cluster, while others arise on the boundary of the small
cluster. The boundaries of the various clusters typically contain two 
types of agents. The red and green nodes in (b) represent two types of
agents.}
\label{fig:structure}
\end{figure}

\begin{table}
\centering
\begin{tabular}{|c|c|c|c|c|c|c|}
\hline
Index of structure &1 &2 &3 &4 &5 &6\\\hline
1  &S1 &S1 &S1 &S1 &S1 &S1\\\hline
2  &S1 &S1 &S1 &S1 &S1 &S2\\\hline
3  &S1 &S1 &S1 &S1 &S1 &S3\\\hline
4  &S1 &S1 &S1 &S1 &S2 &S1\\\hline
\dots &\dots &\dots &\dots &\dots &\dots &\dots\\\hline
729 &S3 &S3 &S3 &S3 &S3 &S3\\\hline
\end{tabular}
\caption{Indexing scheme for the six neighbor structures}
\end{table}

\begin{figure*}
\centering
\includegraphics[width=\linewidth]{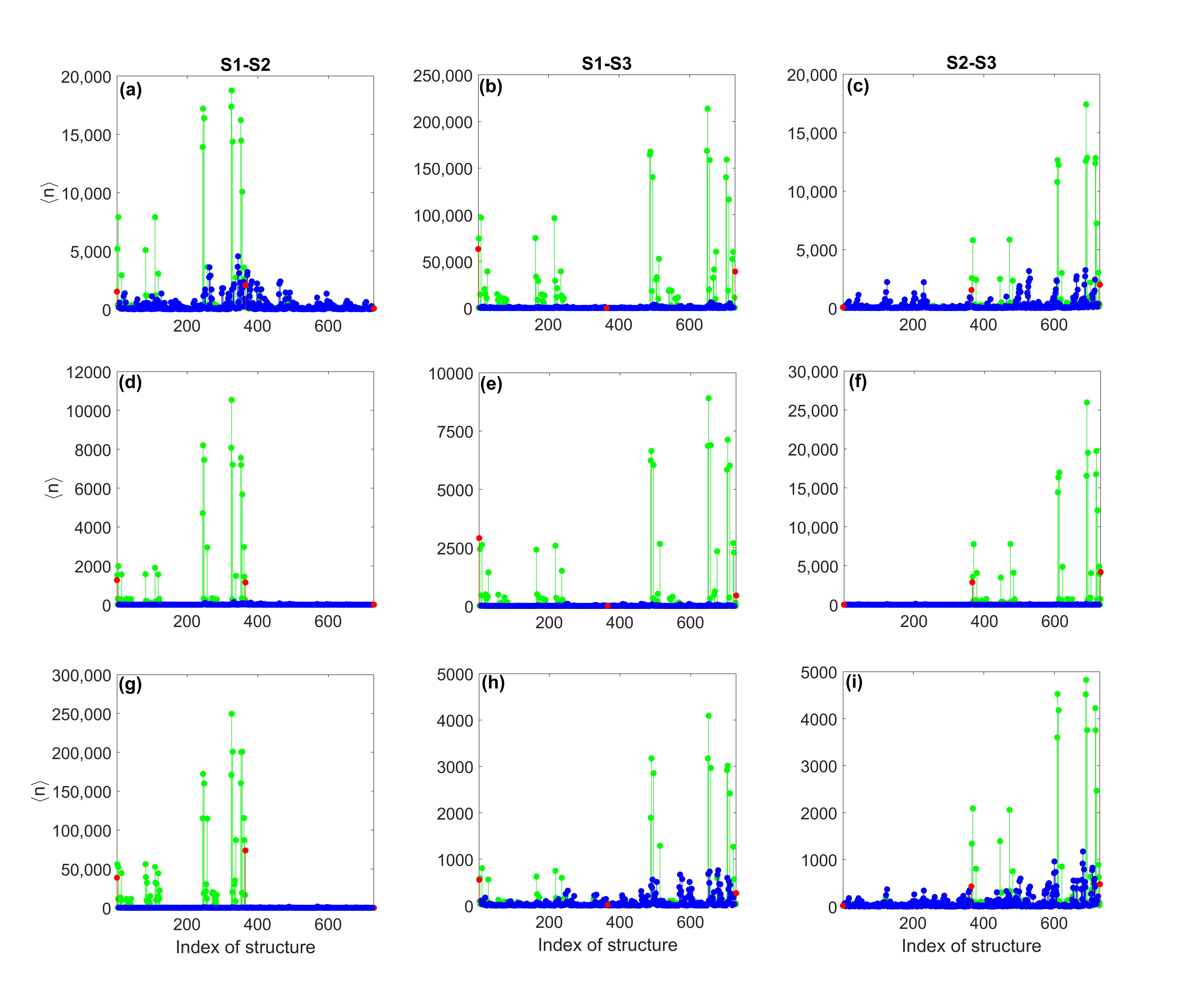}
\caption{ (Color online) {\em Statistical distribution of distinct strategy 
configurations.} (a-i) Counts of every neighboring structure from simulation 
data, where the red (medium gray), green (light gray), and blue (dark gray) 
dots represent the cases where there is only one type, two
types, and three types of agents within the six neighbors 
[Fig.~\ref{fig:structure}(a)], respectively. The values of the parameter $\rho$
are: $\rho=0.01$ for (a-c), $\rho=0.35$ for (d-f), and $\rho=0.99$ for (g-i). 
The distinct boundary structures are: target agent $S^{(1)}$ versus 
opponent agent $S^{(2)}$ (a,d,g), $S^{(1)}$ versus $S^{(3)}$ (b,e,h), and
$S^{(2)}$ versus $S^{(3)}$ (c,f,i). For each panel, the structure index 
is an encoding of the six neighbors in Fig.~\ref{fig:structure}(a), where
are 729 such combinations. All data points are the result of taking the 
average between time steps 251 and 300. The lattice size is $400\times400$.} 
\label{fig:BS_count}
\end{figure*}

\subsection{Modeling of interaction configurations and pair approximation 
on interaction motifs}

Methods of theoretical analysis of the evolutionary dynamics on square 
include the mean-field theory~\cite{SVS:2005} in combination with pair 
approximation~\cite{WW:2007,PM:2006,HS:2005} and the master 
equation~\cite{Gleeson:2013}. To develop a theoretical understanding of
complex dynamical behaviors as exemplified in 
Figs.~\ref{fig:allrho}-\ref{fig:Pattern1}, we take the mean field/pair 
approximation approach. To enumerate the possible pairwise interactions,
we assume that each strategy exists in the clusters of the agents adopting 
that strategy (Fig.~\ref{fig:Pattern1}), so the interactions (strategy 
transitions) occur only at the boundaries of the clusters between different 
strategies. Figure~\ref{fig:structure}(a) shows the typical configurations 
of pair interactions on a square lattice, where the two focal nodes are 
surrounded by six other nodes, and each of the eight nodes can adopt 
any of the three strategies. However, the boundaries between the clusters 
of different strategies typically contain two distinct strategies only, 
which can be empirically verified via the statistics of the strategy 
distribution configurations of the eight-node motif.

For two clusters with regularly shaped boundaries, there are altogether 
six typical configurations of strategies on the eight-node motif, as shown 
in Fig.~\ref{fig:structure}(b). The focal node on the left and its left 
neighbor can be assumed to have the same strategy, as they belong to the
same cluster. The upper and lower neighbors of the focal node can have the
same strategy or the strategy of the other focal node on the right, which 
is different from that of the left focal node. Similarly, the focal node 
on the right and its right neighbor are in the same cluster and thus have 
the same strategy, while its upper and lower neighbor can choose to have 
either of the two strategies freely. Due to the fact that the payoff of 
each focal node depends only on the number of its neighbors in each type 
of two strategies, symmetric configurations are regarded as the same. 
Consequently, there are only six distinct cases, which are denoted as
configurations (1-6) in Fig.~\ref{fig:structure}(b). For boundaries with 
an irregular shape, the left (or right) neighbor of the left (or right) 
focal node on the motif may have a different strategy. Accordingly, there
are three additional configurations, denoted as cases (7-9) in 
Fig.~\ref{fig:structure}(b). Given two distinct strategies, $S^{(x)}$ 
and $S^{(y)}$, the payoffs of the two focal nodes in each of the nine 
configurations can be calculated. Furthermore, under the assumption that 
the nine configurations occur with equal probability, the average payoff of 
each focal node can be obtained, leading to the probability 
$P_{S^{(x)}\rightarrow S^{(y)}}$ of a focal node to replaced by its opponent 
focal node, namely, the probability for the strategy of a focal node to be 
excluded by that of the other focal node.

To gain insights, we  
study the statistical distributions of distinct strategy configurations,
with results shown in Fig.~\ref{fig:BS_count}. The coding scheme of the 
six-neighbor structures is shown in Table 1. We see that the green dots 
occur most frequently at the cluster boundaries, indicating that the 
boundaries are mostly between two types of agents. The most frequently 
occurring boundary structures are those shown Fig~\ref{fig:structure}(b). 
From Fig.~\ref{fig:Pattern1}, we see that there are more small clusters 
for $\rho=0.01$ or $\rho=0.99$ than for $\rho=0.35$. The six large cluster 
boundary structures appear more often for $\rho=0.35$. For $\rho=0.01$, 
the frequencies of $S^{(1)}$ and $S^{(3)}$ are larger than that of 
$S^{(2)}$. There are many more boundary points between $S^{(1)}$ and 
$S^{(3)}$ than any other combinations of agent pairs. Based on the two 
types of agent boundaries and the nine kinds of cluster boundaries, we can
exploit the mean field theory (below) to calculate the theoretical transfer 
probability for any type of agents and predict their frequencies.

The various probabilities of strategy adoptions 
can be calculated by resorting to the pairwise interaction approximation.
For each of the nine strategy distribution configurations on 
the eight-node interaction motif in Fig.~\ref{fig:structure}, the payoffs 
of the two focal nodes with strategies $S^{(x)}$ and $S^{(y)}$ are 
$U_x^{(1)} = (u_{xx} + u_{xy}) + 2 u_{xx}$,
$U_y^{(1)} = (u_{yy} + u_{yx}) + 2 u_{yx}$, 
$U_x^{(2)} = (u_{xx} + u_{xy}) + 2 u_{xx}$,
$U_y^{(2)} = (u_{yy} + u_{yx}) + u_{yx} + u_{yy}$,
$U_x^{(3)} = (u_{xx} + u_{xy}) + u_{xx} + u_{xy}$,
$U_y^{(3)} = (u_{yy} + u_{yx}) + u_{yx} + u_{yy}$,
$U_x^{(4)} = (u_{xx} + u_{xy}) + 2 u_{xx}$,
$U_y^{(4)} = (u_{yy} + u_{yx}) + 2 u_{yy}$,
$U_x^{(5)} = (u_{xx} + u_{xy}) + u_{xx} + u_{xy}$,
$U_y^{(5)} = (u_{yy} + u_{yx}) + 2 u_{yy}$,
$U_x^{(6)} = (u_{xx} + u_{xy}) + 2 u_{xy}$,
$U_y^{(6)} = (u_{yy} + u_{yx}) + 2 u_{yy}$,
$U_x^{(7)} = (u_{xx} + u_{xy}) + u_{xx} + u_{xy}$,
$U_y^{(7)} = (u_{yy} + u_{yx}) + 2 u_{yx}$,
$U_x^{(8)} = (u_{xx} + u_{xy}) + 2 u_{xy}$,
$U_y^{(8)} = (u_{yy} + u_{yx}) + u_{yx} + u_{yy}$,
$U_x^{(9)} = (u_{xx} + u_{xy}) + 2 u_{xy}$, and
$U_y^{(9)} = (u_{yy} + u_{yx}) + 2 u_{yx}$,
where $(x,y) = 1,2,3$ but $x \neq y$. The quantities $u_{xx}$, $u_{xy}$, 
$u_{yx}$, and $u_{yy}$ denote the payoffs of one focal node with strategies 
$S^{(x)}$, $S^{(x)}$, $S^{(y)}$, $S^{(y)}$ against an opponent node with 
strategies $S^{(x)}$, $S^{(y)}$, $S^{(x)}$, and $S^{(y)}$, respectively. 
The quantity $U_x^{(i)}$ (or $U_y^{(i)}$) stands for the total payoff 
obtained by the focal node with strategy $S^{(x)}$ (or $S^{(y)}$) in 
configuration $i$ in the games with its four opponents. 
Accordingly, the probability for a focal node with strategy 
$S^{(x)}$ to be replaced by its opponent focal node with strategy $S^{(y)}$ 
can be calculated as
\begin{equation} \label{eq:pairProb}
P_{S^{(x)}\rightarrow S^{(y)}} =\frac{1}{1 +
\exp[(\langle{U_y}\rangle-\langle{U_x}\rangle)/K]},
\end{equation}
where $\langle{U_x}\rangle = 1/9 \cdot \sum_{i=1}^9 U_x^{(i)}$ and 
$\langle{U_y}\rangle = 1/9 \cdot \sum_{i=1}^9 U_y^{(i)}$. 

\begin{figure}
\centering
\includegraphics[width=\linewidth]{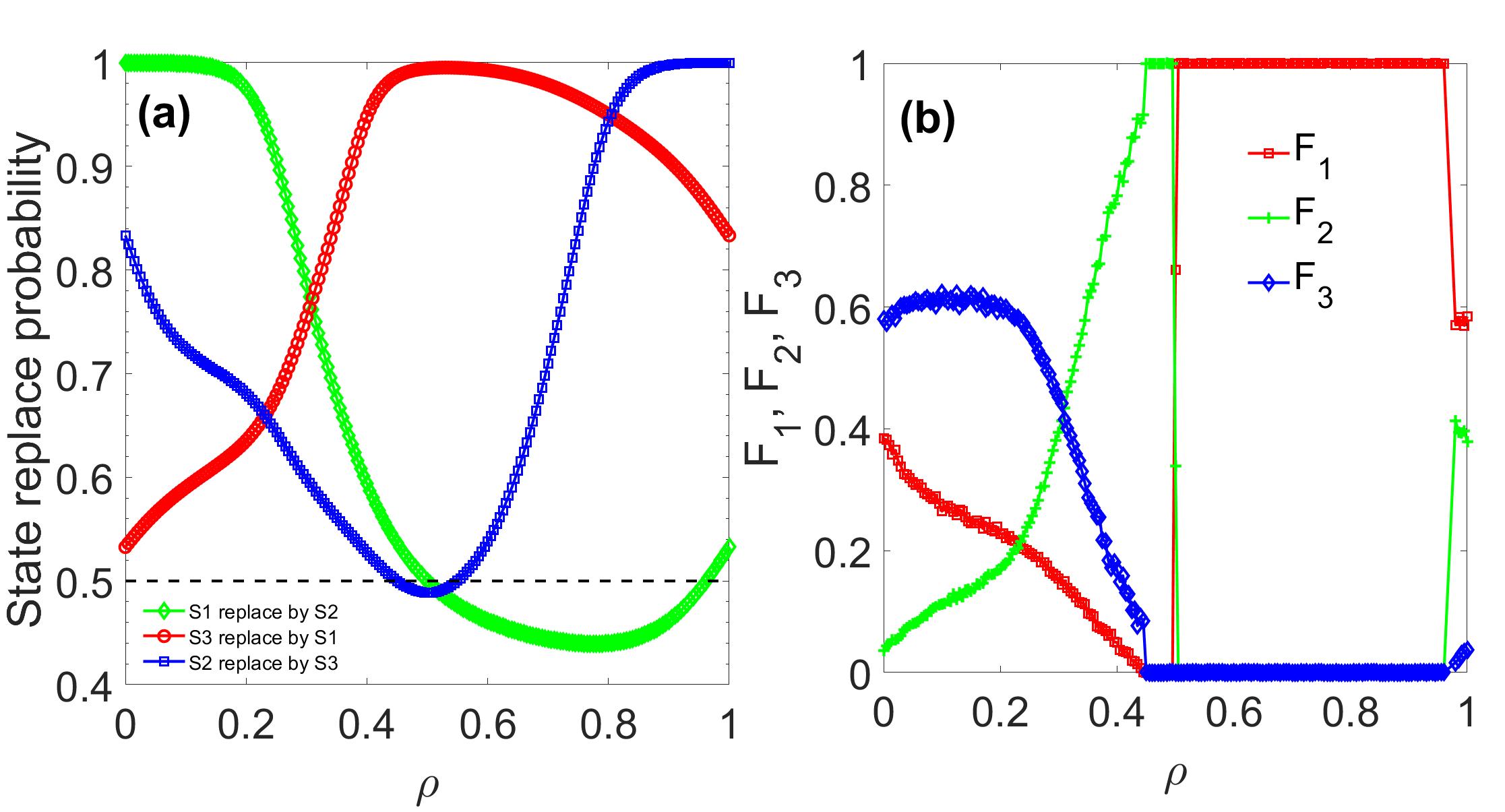}
\caption{ (Color online) {\em Predictions of mean field/pair approximation 
theory.} The replacement probability between any two types of agents and 
the frequency of every type of agents as predicted theoretically, where
the latter is calculated from the mean field theory based on the former.
(a) Theoretically predicted replacement probability based on the nine 
boundary structures in Fig.~\ref{fig:structure}(b). Each point is the 
average value of the replacement probability from the nine boundary 
configurations. The green (light gray), red (medium gray), and blue 
(dark gray) curves are the probabilities of state $S^{(1)}$ replaced 
by $S^{(2)}$, of $S^{(3)}$ replaced by $S^{(1)}$, and of $S^{(2)}$ replaced 
by $S^{(3)}$, respectively. (b) The mean-field predicted frequency of every 
types of agents, where the red (medium gray), green (light gray), and 
blue (dark gray) curves are the frequencies of $S^{(1)}$, $S^{(2)}$, and 
$S^{(3)}$ type of agents, respectively, and every point is the average value 
of the last $500$ time steps of total $10000$ mean-field simulation steps. The 
incremental step in $\rho$ is $0.005$.}   
\label{fig:Tfrequency}
\end{figure}

The pairwise interaction based picture suggests that the 
replacement probability $P_{S^{(x)}\rightarrow S^{(y)}}$ can in fact be 
regarded as an approximation of the strategy transformation probability 
on the square lattice.
Figure~\ref{fig:Tfrequency}(a) shows the interdependence between 
$P_{S^{(x)}\rightarrow S^{(y)}}$ and $\rho$, where $x,y \in [1,2,3]$. 
We see that, for $\rho<0.4$ (regions 1, 2, and 3), the values of 
$P_{S^{(1)}\rightarrow S^{(2)}}$, $P_{S^{(3)}\rightarrow S^{(1)}}$, 
and $P_{S^{(2)}\rightarrow S^{(3)}}$ are all above $0.5$, indicating 
the following RPS mechanism: $S^{(2)}$ excludes $S^{(1)}$, $S^{(1)}$ 
drives out $S^{(3)}$, and $S^{(3)}$ precludes $S^{(2)}$. For 
$0.495 < \rho < 0.96$, we have $P_{S^{(1)}\rightarrow S^{(2)}} < 0.5$, 
which means $P_{S^{(2)}\rightarrow S^{(1)}} > 0.5$, so $S^{(1)}$ actually 
ousts $S^{(2)}$ in this case. Since $P_{S^{(3)}\rightarrow S^{(1)}} \approx 1$,
$S^{(1)}$ eliminates both $S^{(2)}$ and $S^{(3)}$, and this explains the 
dominance of $S^{(1)}$ in region 4. Our pair approximation method 
thus provides an understanding of the qualitative behavior of the 
system in a wide parameter range through local interactions.

\subsection{Mean-field theory}

From a pairwise interaction based microscopic analysis of the 
replacement probability, the behaviors of the hypergame dynamics on a global 
scale can be understood quantitatively.
In terms of the mean field theory, the frequencies of the three strategies 
in the system are governed by the following master equation
\begin{eqnarray}
\frac{dF_1}{dt} &=& [{P_{S^{(2)}\rightarrow S^{(1)}}}-{P_{S^{(1)}\rightarrow S^{(2)}}}]\cdot F_1F_2 \\\nonumber 
& + & [{P_{S^{(3)}\rightarrow S^{(1)}}}-{P_{S^{(1)}\rightarrow S^{(3)}}}]\cdot F_1F_3,\\\nonumber
\frac{dF_2}{dt}&=&[{P_{S^{(1)}\rightarrow S^{(2)}}}-{P_{S^{(2)}\rightarrow S^{(1)}}}]\cdot F_1F_2 \\\nonumber
& + & [{P_{S^{(3)}\rightarrow S^{(2)}}}-{P_{S^{(2)}\rightarrow S^{(3)}}}]\cdot F_2F_3,\\\nonumber
F_3&=& 1-F_1-F_2,
\end{eqnarray}
where $F_1$, $F_2$, $F_3$ are the frequencies of $S^{(1)}$, $S^{(2)}$, 
and $S^{(3)}$, respectively. 
The numerical solution of the master equation group gives estimates of 
the frequencies of the three strategies.
A comparison between Figs.~\ref{fig:allrho} and \ref{fig:Tfrequency}(b) 
reveals that our mean-field calculation captures the essential dynamical
behavior of the system in a wide range of parameter $\rho$ (regions 2 and 4 
as well as the right side of region 1). For $\rho<0.445$, the behaviors 
of $F_1$, $F_2$, and $F_3$ from the mean-field theory coincide with the 
behavior of the real system in region 2, the RPS-like state, almost exactly. 
For $0.445 < \rho < 0.495$, the dominant behavior of $S^{(2)}$ is similar 
to that in the ending part of region 3. For $0.495 < \rho < 0.96$, the 
dominance of $S^{(1)}$ is well reproduced by the theory. For $\rho > 0.96$, 
the numerically observed advantage of $S^{(2)}$ is reproduced. The 
transition points between the various regions are determined by the 
points where the probabilities $P_{S^{(x)}\rightarrow S^{(y)}}$ cross 
the line of $P_{S^{(x)}\rightarrow S^{(y)}}=0.5$ in 
Fig.~\ref{fig:Tfrequency}(a) as $\rho$ increases, due to the flip over
of the ``predator-prey'' relation between $S^{(x)}$ and $S^{(y)}$.

Our mean field theory fails to predict the behaviors in region 3, where 
the equilibrium state can be the absorption state of two strategies. 
There are two reasons for this: extremely long transient time and finite
size effect. Firstly, to predict which strategy would take over 
according to the evolution pattern even after a very long simulation time 
(a typical transient time is $0.5\times 10^6$ time steps) is not feasible, 
since it is not possible to determine which strategy would die first when 
the number of critical time steps determining the fate of a strategy is 
negligibly small. Our numerical simulations show that, even if 
the frequency of one strategy, $S^{(1)}$ or $S^{(3)}$, approaches zero, 
revival at later time can occur with a high probability. Toward the 
right end of region 2, $F_1$ and $F_3$ are close to zero and, as $\rho$ is 
increased, there comes a point at which one of frequencies actually becomes 
zero so that the system enters into region 3. Secondly, calculations with 
different lattice sizes reveal that the absorption state emerges earlier 
in smaller lattices, and the critical $\rho$ value separating regions 2 
and 3 increases with the lattice size. This implies that a finite-size 
effect plays the role of eliminating $S^{(1)}$ or $S^{(3)}$ to generate 
the one-strategy dominant state. Fluctuation effects are more severe in 
smaller lattices, so $S^{(1)}$ may become extinct or $S^{(1)}$ and $S^{(3)}$ 
clusters may be separated spatially with a higher probability.

\section{Discussion} \label{sec:discussion}

In most existing works on evolutionary game dynamics on 
networks, a basic assumption is that the set of possible strategies is 
common to all players in the 
system~\cite{NM:1992,NBM:1994,OHLN:2006,TDW:2007,DHD:2014,ALCFMYN:2017}.
This assumption is reasonable for a variety of real-world phenomena and,
in certain cases, makes possible a deep mathematical understanding of the 
game dynamics. The consideration motivating our work is that, in the real
world, there can be situations where this assumption is not accurate. For example,
in a social network, the individuals can have different backgrounds
of knowledge, financial status, and experience. It is then conceivable 
that the strategy sets available to different players may not be identical. To 
investigate hypergame dynamics on networks is technically quite difficult, 
and such studies are still rare in the literature. We are led to
consider the simplest setting of three available strategies with any player's
access to two (restrictively mixed-strategies). 
Even for the relatively simple setting, a mathematical treatment is not 
feasible: we thus rely on a combination of numerical computations and 
physical reasoning based on the pairwise interaction and mean field 
approximations to gain insights into evolutionary hypergame dynamics 
on regular networks.
  
We find a variety of dynamical behaviors and equilibrium states, including 
states in which most players use one strategy (one dominant strategy) and 
those where players use multiple strategies (coexisting strategies).
There are parameter regions in which the equilibrium frequencies of the 
strategies are completely predictable (e.g., regions 1, 2, and 4 in 
Fig.~\ref{fig:transient}), but there is also a region of unpredictability
(region 3). We also uncover equilibrium
states characteristic of those from cyclic competition dynamics, e.g., 
RPS-like states. A striking phenomenon is that, a nearly extinct strategy
can revive and dominate the whole system. Qualitatively, this may be 
understood as a consequence of the unpredictability: in a parameter region 
where prediction of the system's asymptotic state is ruled out, there can 
be transitions from the RPS-like state. In particular, starting from an RPS 
state, if the advantage of one strategy keeps growing and wins more and 
more agents, the living space of the other two strategies would be 
suppressed. At a certain point, the coverage of the two weak strategies 
would be so low that random fluctuations would remove one of them. If the 
remaining strategy is a prey of the strong strategy, the system would be 
dominated by the strong one. However, if the remaining strategy of the 
two weak ones is the prey of the extinct one, then regardless of its 
weakness, it would eventually overturn the entire population. A 
quantitative understanding of this phenomenon is lacking at the present.

We also find that, in hypergame of the prisoner's dilemma type, 
self-organization of cooperation can be promoted. For example, as the
parameter $\rho$ is increased, the probability of cooperation can 
increase monotonically and reaches the value of close to unity
(Fig.~\ref{fig:Actual_CDL}). Comparing with the traditional prisoner’s 
dilemma game with loneliness~\cite{SH:2002}, in our hypergame, it is not 
necessary for voluntary participation to create a cyclic dominance of 
strategies to promote cooperation.

Our work demonstrates that the diversity in the individuals' understanding 
of the environmental strategies can play an important role in the evolution 
of strategy distribution on a global scale, and it can generate behaviors 
that are fundamentally different from those from the traditional 
explicit-strategy game dynamics. The basic parameter $\rho$ in our model, 
the probability of adopting a strategy, is key to generating the various 
complex dynamical behaviors. This parameter in fact measures the fraction of 
each pure strategy within the restrictively mixed strategy, whose changes 
drive the system into dramatically different equilibrium states. Our study 
reveals that a slight change in the fraction may completely overturn the 
relative advantage between the strategies, suggesting that the game dynamics 
can be manipulated through small changes in the parameter. This opens a door 
to controlling evolutionary hypergame dynamics.  

\section*{Acknowledgement}

We thank Prof.~S.-H. Xu for discussions.
We would like to acknowledge support from the Vannevar Bush Faculty Fellowship 
program sponsored by the Basic Research Office of the Assistant Secretary of
Defense for Research and Engineering and funded by the Office of Naval
Research through Grant No.~N00014-16-1-2828. 

%\renewcommand\refname{Reference}
%\bibliographystyle{apsrev4-1}
%\bibliography{hypergame}

%merlin.mbs apsrev4-1.bst 2010-07-25 4.21a (PWD, AO, DPC) hacked
%Control: key (0)
%Control: author (72) initials jnrlst
%Control: editor formatted (1) identically to author
%Control: production of article title (-1) disabled
%Control: page (0) single
%Control: year (1) truncated
%Control: production of eprint (0) enabled
%

\end{document}